\pdfoutput=1
\documentclass[authoryear, 5p]{elsarticle}

\usepackage[normalem]{ulem}

\usepackage[latin1]{inputenc}       
\usepackage[T1]{fontenc}            
\usepackage{float}
\usepackage[bookmarks,colorlinks, breaklinks]{hyperref}

\usepackage{graphicx}        
\usepackage{multicol}               
\usepackage{multirow}
\usepackage{booktabs}               
\usepackage{algorithm}
\usepackage{algorithmic}
\usepackage[page]{appendix}

\begin{document}
\begin{frontmatter}
\title{\small The final version of this paper has been published in Biological Conservation, 2009, 142, 779-788, doi:10.1016/j.biocon.2008.12.014\\[1cm] \LARGE Smart spatial incentives for market-based conservation \\[0.2cm]}
\journal{Biological Conservation}
\author[UFZ]{Florian Hartig\corref{cor}}
\cortext[cor]{Corresponding author, Tel: +49-341-235-1716, Fax: +49-341-235-1473, \href{http://www.ufz.de/index.php?de=10623}{http://www.ufz.de/index.php?de=10623}}
\ead{florian.hartig@ufz.de}
\author[UFZ]{Martin Drechsler}
\ead{martin.drechsler@ufz.de}

\address[UFZ]{UFZ - Helmholtz Centre for Environmental Research, Department of Ecological Modelling, Permoserstr. 15, 04318 Leipzig, Germany}
\begin{abstract}
Market-based instruments such as payments, auctions or tradable permits have been proposed as flexible and cost-effective instruments for biodiversity conservation on private lands. Trading the service of conservation requires one to define a metric that determines the extent to which a conserved site adds to the regional conservation objective. Yet, while markets for conservation are widely discussed and increasingly applied, little research has been conducted on explicitly accounting for spatial ecological processes in the trading. In this paper, we use a coupled ecological-economic simulation model to examine how spatial connectivity may be considered in the financial incentives created by a market-based conservation scheme. Land use decisions, driven by changing conservation costs and the conservation market, are simulated by an agent-based model of land users. On top of that, a metapopulation model evaluates the conservational success of the market. We find that optimal spatial incentives for agents correlate with species characteristics such as the dispersal distance, but they also depend on the spatio-temporal distribution of conservation costs. We conclude that a combined analysis of ecological and socio-economic conditions should be applied when designing market instruments to protect biodiversity. 
\end{abstract}

\end{frontmatter}

\section{Introduction} 

Market-based instruments such as payments \citep{WUNDER-EfficiencyofPayments-2007, Drechsler-model-basedapproachdesigning-2007}, auctions \citep{Latacz-Lohmann-Auctionsasmeans-1998} or biodiversity offset trading \citep{Panayotou-Conservationofbiodiversity-1994, Chomitz-Transferabledevelopmentrights-2004} have been suggested as a means to complement existing reserves by inducing biodiversity protection on private lands. Market-based instruments are currently being used or tested in many countries around the world. Some examples are conservation and wetland mitigation banking in the US \citep{Salzman-Currenciesandcommodification-2000, Wilcove-Usingeconomicand-2004, Fox-Statusofspecies-2005} or markets schemes in Australia \citep{Coggan-MarketBasedInstruments-2005, Latacz-Lohmann-AuctionsConservationContracts-2005}. One of the reasons for the increasing popularity of these instruments is the realization that markets may achieve a more targeted and therefore more cost-efficient correction of a conservation problem, in particular because landowners have more information about their local costs and can choose the allocation of conservation measures accordingly \citep{Jack-Designingpaymentsecosystem-2008}. Another reason is that market-based instruments are well suited for targeting multiple ecosystem services (e.g. conservation and carbon sequestration \citep{Nelson-Efficiencyofincentives-2008}), a point which has been highlighted in a recent statement of the European Union \citep{EU-Commision-Greenpapermarket-based-2007}.\\\\
At the same time, however, there has been considerable concern over whether current implementations of conservation markets target the right entities. At present, market-based policies for conservation tend to use simple and indirect incentives, such as payments for certain farming practices \citep{Ferraro-DirectPaymentsto-2002}. But are those incentives efficient in protecting threatened species, or are we paying "money for nothing" \citep{Ferraro-Moneynothingcall-2006}? Examining the structure of the given incentives for landowners is the key to answering these questions. What defines a unit of conservation? What are we paying landowners for?\\\\
The overall goal of global conservation efforts is to ensure the persistence of biodiversity in our landscapes \citep{Margules-Systematicconservationplanning-2000}. Therefore, it would be ideal to assess the market value of a conservation measure directly by assessing its effect on species survival \citep{Williams-Usingprobabilityof-2000, Bruggeman-ShouldHabitatTrading-2008}. Unfortunately, applying this method to real-world situations is often not feasible because direct monitoring or detailed population models are too expensive or not available \citep{Jack-Designingpaymentsecosystem-2008}. Moreover, the efficiency of markets crucially depends on the information available to landowners. If landowners do not understand the evaluation criteria for their land, they may choose suboptimal land configurations, or they may decide not to participate in the market at all. Therefore, practically all existing market schemes use a metric, given by a number of indices, that relates measurable quantities of a site (e.g. size) to the site's market value.\\\\
Most of these existing schemes (e.g. habitat banking in the US) base their evaluation solely on the quality and size of the local site without considering its surroundings. This raises some concern because in many cases, the ecological value of a typical private property (e.g. arable field, forest lot) does in fact depend on neighboring properties. Populations or ecosystems may exhibit thresholds for the effectiveness of conservation measures, which implies that a local measure may be ineffective when it is not accompanied by other measures. \citep{Hanski-Minimumviablemetapopulation-1996, Scheffer-Catastrophicshiftsin-2001}. Furthermore, for many endangered species, not only the absolute loss of habitat area, but also habitat fragmentation is a major cause of population decline \citep[compare e.g.][]{Saunders-BiologicalConsequencesOf-1991, Fahrig-Effectofhabitat-2002}. Therefore, metrics that only evaluate sites locally may set the wrong incentives because they do not correspond to the real conservational value of a site.\\\\
Spatial metrics that consider the surrounding of a site are available and are widely used for systematic reserve site selection \citep[e.g.][]{Moilanen-ReserveSelectionUsing-2005, Teeffelen-ConnectivityProbabilitiesand-2006}. Yet, simply transferring spatial metrics from conservation planning into connectivity dependent incentives for landowners (in the following we will call such incentives short "spatial incentives") would be short sighted. Conservation planning metrics have been developed for assessing and optimizing the ecological value of a habitat network from the viewpoint of a planner who considers the whole landscape. Landowners in conservation markets, on the other hand, react to the given incentives independently and with limited knowledge, striving for maximization of their individual utility rather than maximizing global welfare. The fact that the value of a site depends on neighboring sites implies that land use decisions may create costs or benefits for neighboring landowners. In economics, such costs or benefits are referred to as externalities. It is well known that markets may fail to deliver an optimal allocation of land use in the presence of such externalities \citep{Mills-Transferabledevelopmentrights-1980}. Another problem is that, unless we assume perfect information and unlimited intellectual capacities, we must take into account that landowners may fail to find the optimal adoption of their land use in the presence of complicated spatial evaluation rules \citep{Hartig-Staybythy-2008}. Thus, the need to consider human behavior in metrics for market-based instruments is characterized by a trade-off: Ecological accuracy calls for a metric that is complex enough to capture all details of the relevant ecological processes, but socio-economic reality may suggest compromises towards more practical and robust metrics.\\\\
In this paper, we combine a spatially explicit population model with an agent-based simulation model to assess the effect of connectivity-dependent incentives in a virtual conservation market. One key assumption is that landowners do not react optimally to the given incentives, but base their decisions only on the present land configuration and their estimated costs and benefits for the next period. Thus, we seek to optimize for ecological parameters such as dispersal as well as for economic parameters such as behavior of landowners. To simulate the reactions of landowners towards a given spatial metric, we use the conservation market model introduced in \citep{Hartig-Staybythy-2008}. A spatially explicit metapopulation model is placed on top of the emerging landscape structure to evaluate the conservation success for different species in terms of survival probability at a fixed time horizon.

\section{Methods} \label{sec: methods}
\subsection{Overview and purpose}
\begin{figure}[h]
\centering
\includegraphics [width=6cm]{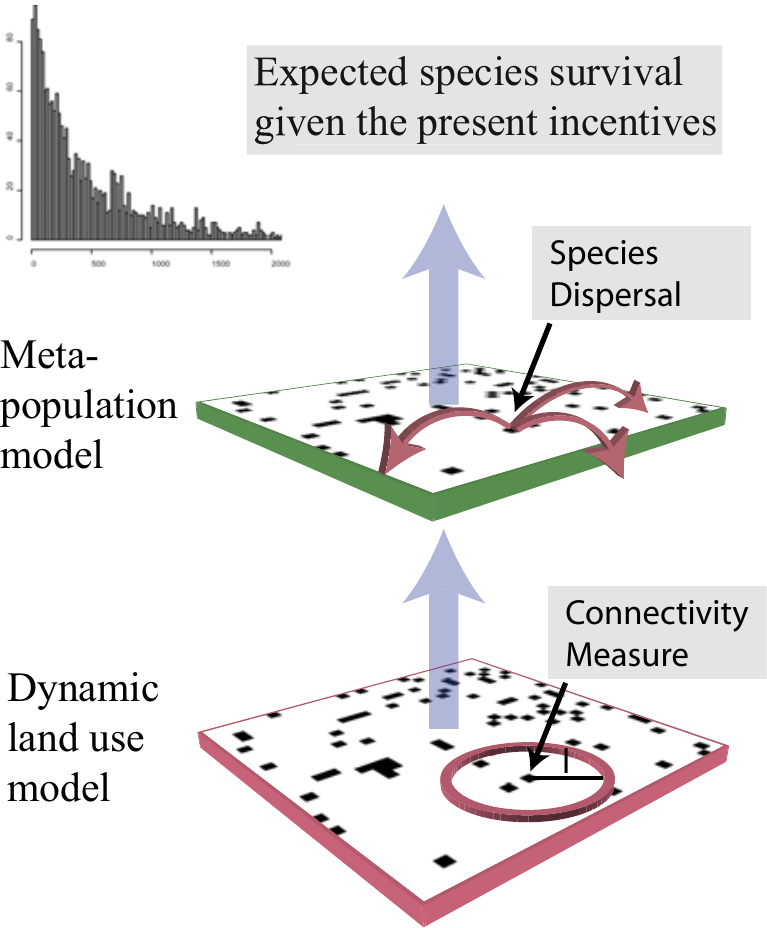}
\caption{Modelling approach: Drivers are spatially heterogeneous, dynamic costs for each site. On the basis of these costs and the spatial incentives, conservation measures are allocated by the economic submodel. The resulting dynamic landscape is used as an input for the ecological model, which estimates species survival probabilities on this landscape.}
\label{figure: modelling approach}
\end{figure}

The aim of this study is to design spatial incentives that result in cost-effective conservation when there are many landowners and the conservation outcome depends on the combination of decisions by landowners. Here, cost-effective means that we maximize the conservation effect at a given budget. The model used contains two submodels: An economic submodel that simulates the trading of conservation credits and an ecological submodel to assess the viability of several species in the dynamic landscape that emerges from the trading activity. The driver for trading and the subsequent change of the landscape configuration is economic change in the region, reflected by heterogeneously changing costs of maintaining a local site in a conserved state. We first describe the state variables of the model, followed by the economic and the ecological submodel and the coupling of the submodels. The coupled model is then used to find the cost-effective metric by comparing the forecasted species persistence across a range of different parameterizations of the metric. Fig. \ref{figure: modelling approach} shows a graphical representation of our model approach.

\subsection{State variables and scales}
The simulation is conducted on a rectangular $30\times30$ grid with periodic boundary conditions (i.e. the grid has the topology of a torus). The $n=30^2$ grid cells represent both the economic (property) units and the ecological (habitat) units. Although the model may be applied to any spatial and temporal scale, we think of grid cells as being of the size of an average agricultural field in Europe (around $10$ ha), and time steps being a year. Grid cells $x_i$ occur in two states: They can be conserved at a cost $c_i$ and thus provide habitat for the species, or they are used for other economic purposes, resulting in no costs. The conservation state of a grid cell is labelled with $\sigma_i$, $\sigma_i = 1$ being a conserved cell and $\sigma_i=0$ being an unconserved cell. Conserved grid cells may be either occupied (populated) $p_i = 1$ or unoccupied $p_i = 0$ by the species under consideration. Unconserved grid cells can never be occupied. A list of the state variables and parameters of the two submodels is given in Table \ref{table:state variables}.

\begin{table}[h]
  \centering
  \begin{tabular}{l@{\hspace{0.2cm}}l@{\hspace{0.2cm}}l} \toprule
  \textsc{Symbol} & \textsc{Connotation} & \textsc{Range}  \\ \midrule \addlinespace[0.2cm] 
  \multicolumn{2}{l}{\uline{State variables:}} & \\
  $x_i$ & Position of the i-th cell on the grid &  \\ 
  $\sigma_i(t)$ & Conservation state of the i-th cell  &  $\{0,1\}$ \\
  $p_i(t)$ & Population state of the i-th cell  &  $\{0,1\}$ \\
  $c_i(t)$ & Opportunity costs of $\sigma_i$ = 1 at t&  around $1$  \\[0.2cm] 
  \multicolumn{2}{l}{\uline{Parameters economic model:}} & \\
  $\Delta$ & Cost heterogeneity & $[0..1]$ \\
  $\omega$ & Cost correlation & $[0..\infty]$ \\
  $m$ & Connectivity weight  & $[0..1]$ \\  
  $l$ & Connectivity length  & $[0..\infty]$ \\ 
  $\lambda $ & Budget constraint & $[0..\infty]$\\[0.2cm]
  \multicolumn{2}{l}{\uline{Parameters ecological model:}} & \\
  $e$ & Local extinction risk& $[0..1]$\\
  $r$ & Emigration rate & $[0..\infty]$\\
  $r_d$& Emigration rate after destruction & $[0..\infty]$\\
  $\alpha^{-1} $ & Dispersal distance & $[0..\infty]$\\ \bottomrule \\
\end{tabular}
\caption{List of state variables (top), parameters of the economic model (middle) and parameters of the ecological model (bottom). Note that although we omit to denote the time dependence $(t)$ explicitly throughout the main text, all state variables and expressions derived from state variables are time dependent.} 
\label{table:state variables}
\end{table}

\subsection{Economic model}\label{sec: economic model}
\begin{figure}[h]
\centering
\includegraphics []{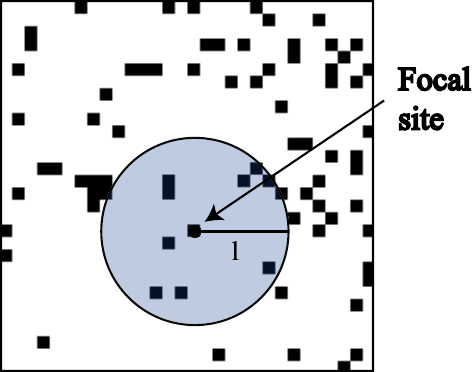}
\caption{Illustration of the connectivity measure: The connectivity $\zeta_i(l)$ is the fraction of conserved sites within a circle of radius $l$ of the focal site}
\label{figure: connecticity measure}
\end{figure}

The economic model describes the decisions of landowners to establish, maintain, or quit a conservation measure on their land (grid cell) in each period. Landowners decisions are based on whether conservation or alternative land use generates a higher return. The returns on the two land use types are influenced by dynamic, spatially heterogeneous costs for conserving a grid cell and by the metric of the conservation market, which decides on the amount of conservation credits to be earned with a particular site, and by the current market price for conservation credits. The model is designed as a spatially explicit, agent-based partial equilibrium model \citep[compare][]{ Drechsler-Applyingtradablepermits-2008, Hartig-Staybythy-2008}.\\\\
A conserved grid cell~$x_i$ produces a certain amount of conservation credits~$\xi_i$ depending on the number of conserved grid cells in its neighborhood. We use the following metric to determine $\xi_i$:
\begin{equation}\label{eq: incentive}
 \xi_i = (1-m) + m \cdot \zeta_i(l) \;.   
\end{equation}
The first term $1-m$ is independent of the connectivity and may be seen as a base reward for the conserved area. The parameter $m$ is a weighting factor that determines the importance of connectivity compared to area. The second term $m \cdot \zeta(l)$ includes the connectivity of the site, measured by the proportion of conserved sites within a circle of radius $l$:
\begin{equation}\label{eq: connectivity measure}
 \zeta_i(l) =  \left( \sum_{d_{ij}<l} \sigma_j \right) \cdot \left( \sum_{d_{ij}<l} 1\right)^{-1} \;.      
\end{equation}
Here, $d_{ij}$ refers to the distance between the focal cell $x_i$ and another cell $x_j$. Fig. \ref{figure: connecticity measure} shows a graphical illustration of this connectivity measure. The total amount of credits in the market is given by the sum of $\xi_i$ over all conserved grid cells:
\begin{equation}\label{eq: total credits}
    U = \sum_{i=1}^n \sigma_i \xi_i \;.
\end{equation}
The conservation of a site results in costs that differ among grid cells. Conservation costs may vary over space and time \citep{Ando-SpeciesDistributionsLand-1998, Polasky-Wheretoput-2008}. We use three different algorithms to generate pattern of random dynamic costs $c_i(t)$. All algorithms create average costs of 1, but they differ in the spatial and temporal distribution of costs. Algorithm \ref{alg cost random} generates spatially and temporally uncorrelated random costs by drawing from a uniform distribution of width $2 \Delta$ at each time step. Algorithm \ref{alg cost random walk} creates spatially uncorrelated, but temporally correlated costs by applying on each grid cell a random walk of maximum step length $\Delta$ together with a small rebounding effect that pushes costs towards $1$ with strength $\omega$. Algorithm \ref{alg cost correlated walk} creates spatio-temporally correlated costs, using a random walk of maximum step length $\Delta$ combined with a spatial correlation term that pushes costs with strength $\omega$ towards the average costs in the neighborhood. A mathematical description of the three algorithms is given in appendix \ref{appendix: Cost algorithms}, together with figures of the created cost distributions (Fig. \ref{figure: cost surfaces} and Fig. \ref{figure: cost time series}).\\\\
To simulate trading, we introduced a market price $P$ for credits. The benefits to be earned by a site are given by $P\cdot \xi$ where $\xi$ is the amount of credits to be earned by a site (eq.~\ref{eq: incentive}). Based on his costs and the potential benefits, each landowner decides whether to conserve his land or not. The model has two options for determining the equilibrium price of the market: Either the price is adjusted until a certain target level for the total amount of produced conservation credits $U$ (eq. \ref{eq: total credits}) is met, or the price is adjusted until a certain level of aggregated costs for the conservation is reached. By aggregated costs, we mean the sum of the costs of all conserved sites:
\begin{equation}\label{eq: aggregated costs}
    C = \sum_{i=1}^n \sigma_i c_i \;.
\end{equation}
Fixing the target reflects a situation where the quantity of conservation credits is fixed. This is, for example, the case in a tradable permit scheme. Fixing the costs, on the other hand, could correspond to a payment scheme where a conservation agency buys credits until a budget constraint is reached. The two options differ when global properties of the cost distribution, such as the mean, change over time. In our simulation, however, costs are in a steady state that is normalized to a mean of $1$. Thus, both options are approximately identical except for finite size effects, which would disappear in the limit of an infinitely large landscape. We chose the second option of fixing the budget for the analysis because it allows an easier comparison between different metrics. Appendix \ref{appendix: economic scheduling} gives a detailed description of the scheduling of the economic model.\\\\
 
\subsection{Ecological model}
To evaluate conservation success in the emerging dynamic landscapes, we use a stochastic metapopulation model \citep[][]{Hanski-Metapopulationdynamics-1998, Hanski-MetapopulationEcology-1999}. Each conserved grid cell is treated as a habitat patch, meaning that each grid cell may hold a local population of the species. Local populations produce emigrants which may disperse and establish a new local population on an unoccupied cell. At the same time, local populations are subject to local extinction, which may be caused e.g. by demographic or environmental stochasticity. The population as a whole can persist on the landscape if the average recolonization rate is higher than the average local extinction risk, yet, stochastic fluctuations of the number of occupied patches may eventually cause extinction of the whole metapopulation. The better the connectivity among patches, and the more patches in the network, the lower the probability of such a global extinction. \\\\
Local extinctions are modelled by a constant chance $e$ of each local population to go extinct per time step. The amount of dispersers arriving from a source patch $x_j$ at a target patch $x_i$ is given by the following dispersal kernel:
\begin{equation}\label{eq: dispersal kernel}
    p_{ij} = r\cdot \frac{1}{\sum_i \sigma_i - 1} \cdot e^{- \alpha \cdot  d_{ij}}\;,
\end{equation}
where $r$ is the emigration rate, the term $(\sum_i \sigma_i - 1)^{-1}$ divides the number of dispersing individuals by the available habitat patches, and the exponential term describes mortality risk during dispersal as a function of distance between $x_i$ and $x_j$. If a patch has been destroyed at the current time step, we set the emigration rate to $r_d$, assuming that a proportion of $r_d$ of the population will be able to disperse before destruction. The sum of all arriving immigrants according to eq. \ref{eq: dispersal kernel} (truncated to 1) is taken as the probability that this patch is colonized at the current time step. Appendix \ref{appendix: ecologic scheduling} gives a detailed description of the scheduling of the ecological model.

\subsection{Parametrization and analysis of the model}\label{sec: optimization approach}
Different species have different connectivity requirements depending on their dispersal abilities. Therefore, we expect an optimized spatial metric to reflect this by values of the connectivity weight $m$ and the connectivity length $l$ that are related to the species characteristics $r, r_d$ and $\alpha$. Additionally, optimal values for $m$ and $l$ may be affected by economic conditions, i.e. the distribution of conservation costs. To analyze the effect of species characteristics and the cost distribution on the optimal spatial incentive, we varied both the connectivity weight $m$ and the connectivity length $l$ of the metric eq. \ref{eq: incentive} for three different cost scenarios and for three different species types.\\\\
The three cost scenarios were generated by Algorithm~\ref{alg cost random} at $\Delta = 0.2$, Algorithm~\ref{alg cost random walk} at $\Delta = 5 \cdot 10^{-5}$ and $\omega = 0.0065$, and Algorithm \ref{alg cost correlated walk} at $\Delta = 0.015$ and $\omega = 0.006$. Table \ref{table: cost algorithms} displays a summary of the three scenarios. Remember that the first scenario creates uncorrelated costs, the second creates temporally correlated costs and the third scenario creates spatio-temporally correlated costs. Figs. \ref{figure: cost surfaces} and \ref{figure: cost time series} in Appendix \ref{appendix: Cost algorithms} show the spatial and temporal cost distribution generated by the chosen parameters.
\begin{table}[h]
  \centering
  \begin{tabular}{l@{\hspace{0.2cm}}l@{\hspace{0.2cm}}l} \toprule
  \textsc{Cost Scenario} & \textsc{Parameters} &    \textsc{Characteristics}  \\ \midrule \addlinespace[0.2cm] 
  \textsc{\ref{alg cost random} - Random}   & $\Delta = 0.2$ &  uncorrelated \\ 
  \textsc{\ref{alg cost random walk} - Random walk} & $\Delta = 5 \cdot 10^{-5}$&  time correlated \\ 
   & $\omega = 0.0065$ & \\
   \textsc{\ref{alg cost correlated walk} - Correlated walk} &  $\Delta = 0.015$& space and time \\
   & $\omega = 0.006$ &  correlated \\ \bottomrule \\ 
\end{tabular}
\caption{Overview of the cost scenarios created by the three algorithms.}\label{table: cost algorithms}
\end{table}
For the species, we consider three functional types: Short-range, intermediate and global dispersers. The parametrization for the three species is displayed in Table \ref{table: species}. To assess the extinction risk for the species, we ran the simulation with different random economic starting conditions between 300 and 1000 times and calculated the probability of a metapopulation extinction after 1000 time steps.\\\\
\begin{table}[H]
  \centering
  \begin{tabular}{l@{\hspace{0.2cm}}l@{\hspace{0.2cm}}l@{\hspace{0.2cm}}l@{\hspace{0.2cm}}l} \toprule
  \textsc{Species type} & $\mathbf{e}$  &   $\mathbf{r}$  &   $\mathbf{r_d}$ & $\mathbf{\alpha^{-1}}$ \\ \midrule \addlinespace[0.2cm] 
  \textsc{I - short dispersal}   & 0.29  & 3     & 1 & 5 \\ 
  \textsc{II - intermediate dispersal} & 0.51 & 3    &  1  & 25\\ 
 \textsc{III - global dispersal} & 0.66  & 3  & 1   &1000  \\ \bottomrule \\
\end{tabular}
\caption{Parameter values for the three species types considered. $\mathbf{\alpha^{-1}}$ is the typical dispersal distance, measured in units of the grid cell length. With cell lengths of $100$ m, this translates to typical dispersal distances of 0.5 km, 2.5km and 100 km, respectively.} \label{table: species}
\end{table}

The budget constraint $\lambda$ for the aggregated costs (eq. \ref{eq: aggregated costs}) was fixed at 0.03 times the number of grid cells $n$ for scenarios with cost dynamics generated by the random walk algorithms (economic scenarios 2~and~3) and at 0.05 times the number of grid cells $n$ for the scenarios created with the random algorithm (economic scenario~1). Exceptions are the combination economic scenario 3 with species 3, where aggregated costs were set at 0.1 times $n$ and economic scenario 1 with species 3, where aggregated costs were set at 0.18 times $n$. The adjustment to different budgets was done to create similar survival probabilities across the nine scenarios formed by systematic combination of the three cost scenarios and the three species types.

\begin{figure}[h]
\centering
\includegraphics [width=8.2cm]{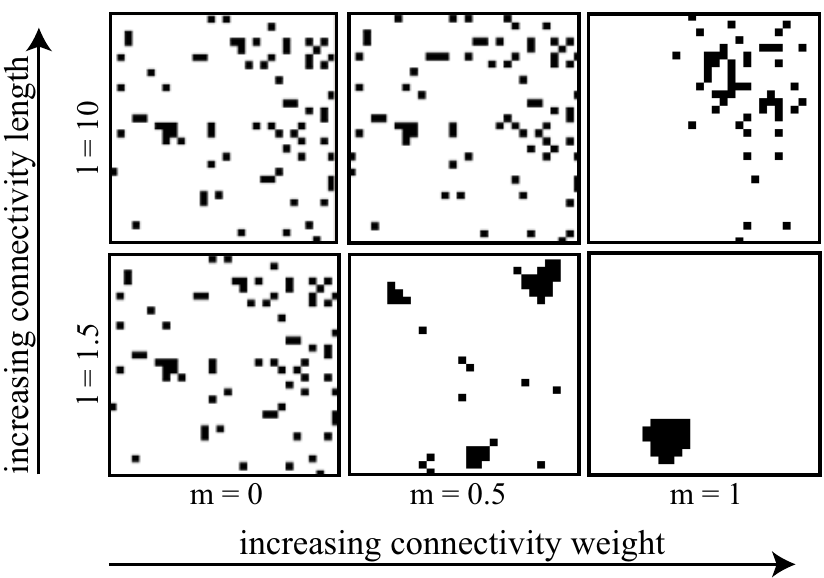}
\caption{Effect of the connectivity weight $m$ and the connectivity length $l$. The pictures show typical landscape structures emerging from trading with costs being sampled by Algorithm \ref{alg cost random walk} at $\Delta = 5 \cdot 10^{-5}, \omega = 0.0065$. Conserved sites are colored black, other sites are colored white. The top row is created with a long connectivity length ($l=10$), the bottom row with a short connectivity length ($l=1.5$). The pictures in the left column are taken at $m=0$, which means that no weight is put on connectivity. Consequently, the landscape structure is dominated by the sites of lowest costs. Increasing connectivity weight ($m=0.5$ middle, $m=1$ right) results in increasing clustering of conserved sites, but in a smaller total area. At a connectivity weight of $m=1$, meaning that all weight is put on connectivity, $l=1.5$ results in a very dense cluster, while the larger connectivity length $l=10$ results in a more spread out configuration.}
\label{figure: landscapes}
\end{figure} 
\section{Results}

\subsection{Emerging landscapes}
For all cost scenarios and all connectivity lengths, an increase in connectivity weight results in more aggregated landscape structures. The density of the clustering is controlled by the connectivity length $l$, which determines how close patches have to be to be counted as connected. Smaller connectivity lengths ($l \sim 1.5$, corresponding to the direct 8-cell neighborhood) result in very dense clusters at full connectivity weight, while larger connectivity lengths lead to more loose agglomerations of conserved sites. Due to the spatial cost heterogeneity, there is a trade-off between clustering and area: At a fixed budget, a higher connectivity weight results in lower total area, but with higher clustering. Typical landscapes are displayed in Fig. \ref{figure: landscapes}.

\begin{figure}[h]
\centering
\includegraphics []{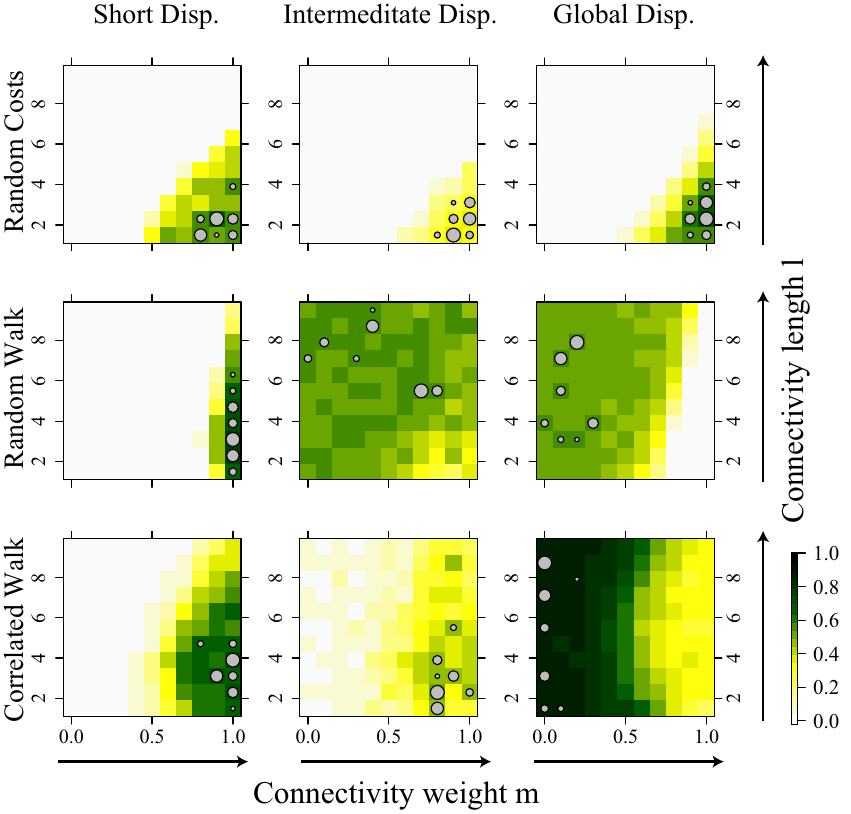}
\caption{Survival probability as a function of  connectivity weight (x-axis) and connectivity length (y-axis) for the three species types (columns 1-3) and for three cost scenarios (rows 1-3). Dark values represent high survival probabilities. The gray circles mark the seven combinations of $m, l$ that yielded the highest survival probabilities, with larger circle size indicating a better ranking within these seven combinations. For most of the scenarios, these optimal points cluster in one small area of the parameter range. The uncertainty of the survival probability can be estimated from a binomial error model. Typical values of the absolute standard error are in the order of $0.01$. This explains why there is some remaining spread of the best combinations of $m, l$ when $m \simeq 0$ is favored (meaning that $l$ has little influence on the model) or when survival probabilities are very similar within a larger area of ($m,l$).}
\label{figure: survival results}
\end{figure}

\subsection{Optimal incentive}
To find the most effective spatial metric ($m, l$), we varied connectivity weight between 0 and 1 and connectivity length between 1.5 and 9.5 in 11 linear steps. Note that a conservation market with no spatial trading rules corresponds to a value of $m=0$. The resulting survival probabilities after 1000 years for the three cost scenarios and the three species types are shown in Fig. \ref{figure: survival results}. The results show that a short disperser such as species I may gain substantially from a very high connectivity weight and short to medium connectivity lengths, while globally dispersing species such as species III benefit from a low connectivity weight and are relatively insensitive towards the connectivity length. For intermediate species such as species II, the tendency changes depending on the cost scenario. An exception is the cost scenario 1 with random costs, which requires very high connectivity weight and short connectivity lengths for all species. We will discuss the  reasons for this in the next subsection. 

\begin{figure}[h]
\centering
\includegraphics []{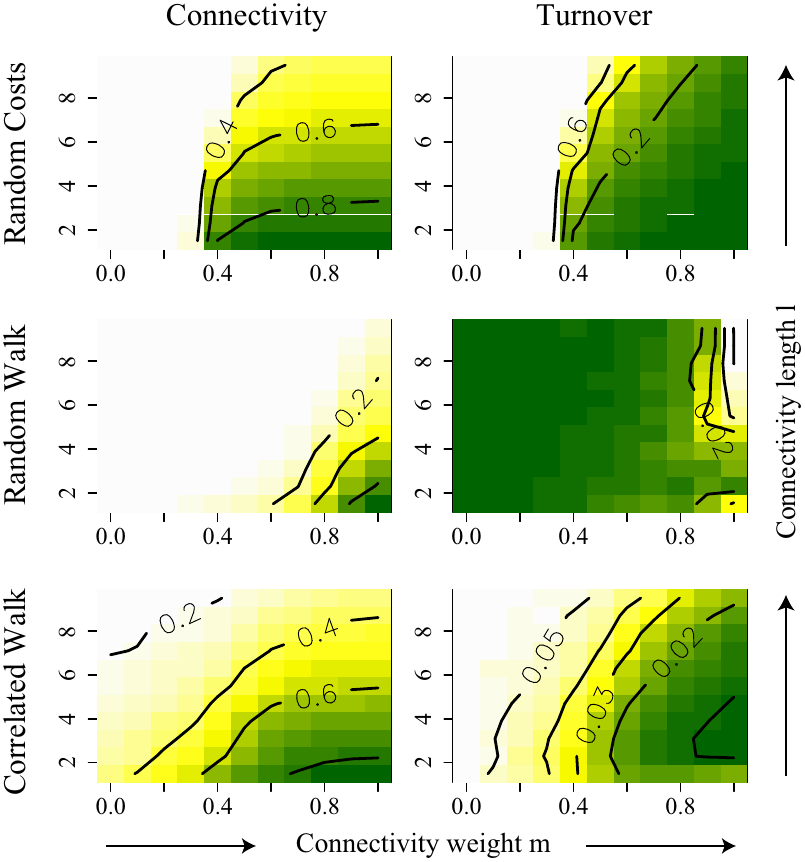}
\caption{Resulting mean connectivity and turnover for the three cost scenarios as a function of connectivity weight $m$ and connectivity length $l$. Connectivity is measured as the mean of $\zeta(1.5)$ of all conserved sites. Turnover, the fraction of conserved sites that are destroyed and recreated elsewhere per time step, serves as an estimate for the intensity of landscape dynamics. Dark values represent low turnover and high connectivity, respectively.}
\label{figure: connectivity and turnover}
\end{figure}

\subsection{Interpretation of the results}
The observed influence of the cost scenarios on the effectiveness of the applied metric ($m, l$) suggests that the emerging landscapes differ among the different cost scenarios. To analyze this difference, we plotted landscape connectivity as well as turnover (the fraction of conserved sites that are destroyed and recreated elsewhere per time step) as a function of the metric parameters $m$ and $l$ for the three considered cost scenarios (Fig. \ref{figure: connectivity and turnover}). \\\\

\begin{figure}[h]
\centering
\includegraphics []{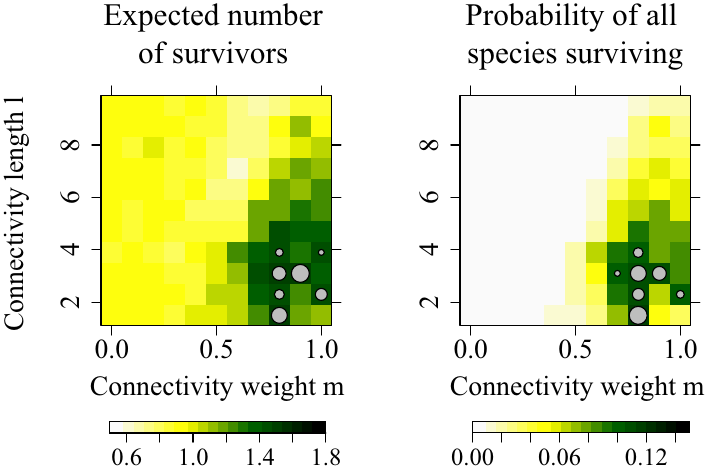}
\caption{Expected number of surviving species (right) and probabilities of all species surviving (left) for the spatio-temporally correlated costs of scenario 3. The gray circles mark the seven combinations of $m, l$ which yielded the highest score of the applied measure, with larger circle size indicating a better ranking within these seven combinations.}
\label{figure: multiple species}
\end{figure}

The results show that greater temporal randomness in the costs causes higher turnover, in that landowners switch rapidly between conserving and not conserving. This increase in turnover effectively increases the local extinction risk, because local populations go extinct at the destroyed sites, while the remaining subpopulations can not immediately recolonize the new sites. Creating connected patches leads to more stability, as neighborhood benefits may outweigh the individual variation in cost for a cell. Thus we are not only facing a trade-off between area and connectivity, but a trade-off between area, connectivity and turnover. The latter explains why different economic scenarios lead to different optimal metrics: For random costs as in scenario~1, turnover rates are very sensitive to the chosen spatial metric. Consequently, turnover totally dominates species survival and high connectivity weight is favored for all species because it reduces the turnover rate. In contrast, the spatial metric hardly affects turnover for scenario~2 and scenario~3. Here, the optimization results (Fig. \ref{figure: survival results}) only reflect the trade-off between connectivity and area: Short range dispersers require high connectivity weights, while global dispersers prefer larger areas.

\subsection{Multiple-species optimization}
Assuming that all three species defined in Table~\ref{table: species} share the same habitat, but do not interact, we can also use our model to generate recommendations on how to support all three species at the same time. There are several options available to combine the survival of multiple species into one index \citep{Nicholson-Objectivesmultiple-speciesconservation-2006, Hartig-timehorizonand-2008}. Here, we use two common indices. The first is the expected number of surviving species, which is given by the sum of the survival probabilities of all species. The second index is the probability of all species surviving, given by the product of the survival probabilities of all species. As for the single species case, both indices were calculated for a time horizon of 1000 years. Fig.~\ref{figure: multiple species} shows the resulting scores for the spatio-temporally correlated cost scenario. Both objectives suggest a moderately strong connectivity weight around $m=0.8$ and a small connectivity length around 3.

\section{Discussion}\label{sec: discussion}
\subsection{Main findings}
We presented a coupled ecological-economic model to optimize spatial incentives in a market for conservation credits. The model shows that conservation markets that consider connectivity lead to considerably better conservation results than markets without spatial incentives (represented by $m=0$ in Fig.\ref{figure: survival results}). Generally, we find that short dispersing species do best with a high weight on connectivity and small-scale connectivity measures. Global dispersers, being largely insensitive to the spatial arrangement of conservation measures, do better with a low weight on connectivity, because this allows the creation of more conserved sites within the given budget. When conserving all species together, a relatively high weight on connectivity yields robustly the highest joint survival probability (Fig \ref{figure: multiple species}). This shows once more that, if connectivity is relevant for the species of concern, spatial evaluation rules may considerably improve the cost-effectiveness of market-based instruments.\\\\
Besides species characteristics, the economic scenarios had an additional, and in some cases large, influence on the optimal spatial metric. The reason is that in the presence of dynamic conservation costs, the spatial incentive does not only influence landscape connectivity, but also landscape dynamics (Fig~\ref{figure: connectivity and turnover}). Landscape dynamics, measured by the rate of turnover (the fraction of conserved sites that are destroyed and recreated elsewhere per time step), negatively affects species survival because the reallocation of a conserved site effectively increases the local extinction risk of the species. In most cases, turnover was negatively correlated with connectivity weight and clustering (Fig~\ref{figure: connectivity and turnover}). The latter explains why under cost scenario~1 (uncorrelated random costs), a stronger connectivity weight is favored for all species: The spatio-temporally uncorrelated costs of this scenario lead to very high turnover rates under a low connectivity weight. Consequently a high connectivity weight that limits the amount of turnover rates is favored for all species.
\subsection{Generality of the results and future research}
The ecological model used for this study neglects a number of factors frequently studied in population models: The landscape is ecologically homogeneous and we have included neither local population dynamics nor a possible dependence of local extinction risk and dispersal on the local population size, nor did we consider correlated environmental stochasticity or catastrophic events. Analyzing the consequences of these factors on the cost-effectiveness of metrics for market-based instruments is a matter of future research. If required all these factors could easily be included without changing the rest of the model, including the analysis method. Furthermore, more sophisticated policies and economic models could be introduced without changing the ecological model.\\\\ 
The main findings of this paper, however, i.e. the positive effect of relatively simple spatial incentives as opposed to no spatial incentives, will qualitatively hold for most realistic scenarios where dispersal is a limiting factor for species. We recommend testing these ideas more often in real-world market schemes such as the examples discussed by \citet{Chomitz-Viablereservenetworks-2006} or \citet{Drechsler-model-basedapproachdesigning-2007}.\\\\
The most apparent shortcoming of the model at this point are simplifications with respect to the time dimension, in particular the inclusion of temporal incentives such as minimum durations of conservation measures on the economic side and time lags for recreation of habitat due to succession on the ecological side. It seems promising for future research to study the control of landscape dynamics through temporal incentives, either independently or in connection with spatial incentives.
\subsection{Consequences for conservation policy}
We believe that our results contain three important messages for conservation policy. The first is that the inclusion of spatial incentives may provide a substantial efficiency gain for conservation markets when fragmentation is a crucial factor for the populations under consideration. Our simulations show that it is possible to account for complicated spatial ecological and economic interactions with relatively simple spatial incentives. Given that most existing market-based conservation schemes worldwide do not explicitly account for spatial processes, it seems promising to examine the potential efficiency gains that could be realized by applying spatially explicit metrics for market-based conservation.\\\\
The second message is that market-based instruments are likely to produce dynamic landscapes, because a voluntary market is based on the possibility that landowners withdraw from conservation measures while others step in for them. This is not a problem in itself. A moderate amount of landscape dynamics may sometimes even benefit the conservation objective. Yet, landscape dynamics must be considered in the design of marked-based instruments and in underlying ecological models. Neglecting dynamics may lead to severe problems for the ecological effectiveness of a market scheme.\\\\
The third message is that optimal spatial incentives are not context-free. The effectiveness of a spatial metric may be sensitive to the economic situation to which it is applied. Thus, a thorough examination of both the ecological as well as the economic and social background is required before deciding on spatial incentives for market-based instruments.

\section{Acknowledgements}
The authors would like to thank Silvia Wissel and Karin Johst for helpful comments during the preparation of the manuscript and Anne Carney for proofreading the text. We are very grateful for the comments which were raised by Doug Bruggeman and two other anonymous reviewers. Their ideas and suggestions greatly contributed to the final manuscript.
 


\begin{appendices}

\renewcommand{\theequation}{A.\arabic{equation}} 
\setcounter{equation}{0} 

\section{Cost algorithms}\label{appendix: Cost algorithms}
\begin{figure}[h]
\centering
\includegraphics[]{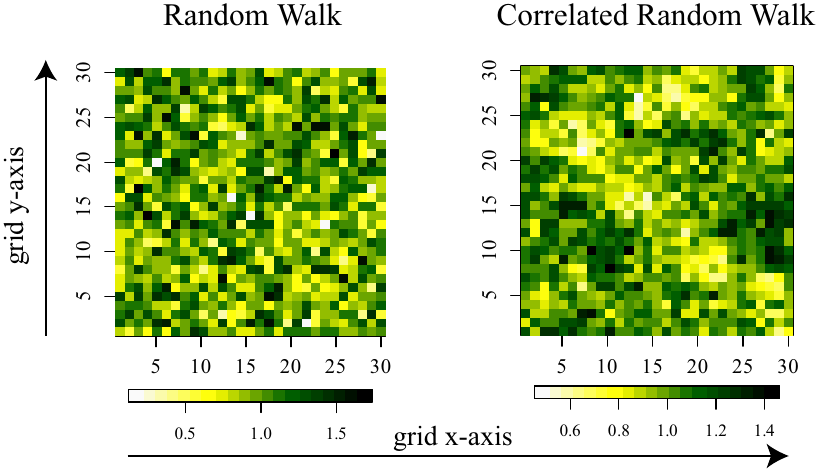}
\caption{Spatial cost distributions generated by the random walk algorithms (Alg. \ref{alg cost random walk} and \ref{alg cost correlated walk}). The two figures shows the $30 \times 30$ grid cells with high cost cells in light and low cost cells in dark colors. The left figure was created by the random walk (algorithm \ref{alg cost random walk}) at $\Delta = 5 \cdot 10^{-5}, \omega = 0.0065$, to the right the correlated random walk (algorithm \ref{alg cost correlated walk}) at $\Delta = 0.015, \omega = 0.006$. Note that low and high cost areas are clustered for the correlated random walk.}
\label{figure: cost surfaces}
\end{figure}

\begin{figure}[h]
\centering
\includegraphics [width=7cm]{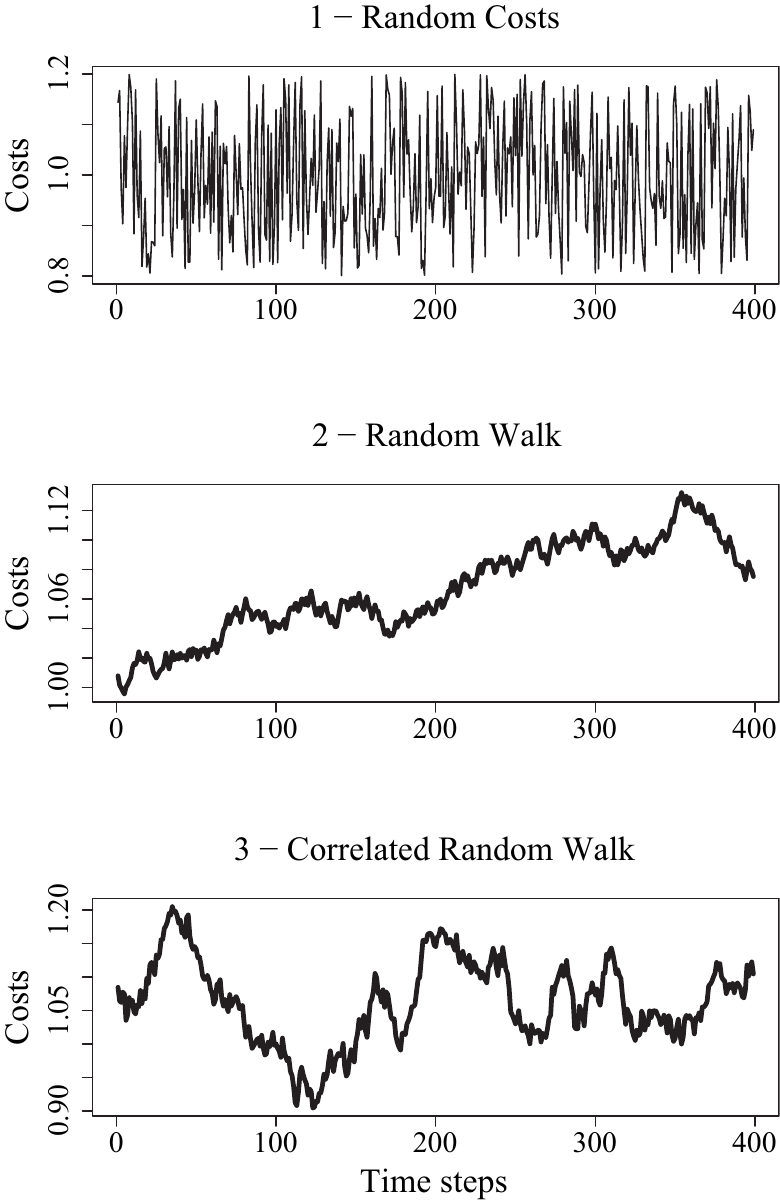}
\caption{Time series of the costs of a grid cell over time. Algorithm \ref{alg cost random} which changes costs randomly at each time step creates a strongly fluctuating time series. The two random walk algorithms lead to a time-correlated series.}
\label{figure: cost time series}
\end{figure}

Alg. \ref{alg cost random} creates random, spatially and temporally uncorrelated costs by drawing the costs of each cell for each time step from a uniform distribution of width $2 \Delta$. The scheduling within one time step is as follows:
\begin{algorithm}[H]
\caption{Random Costs}\label{alg cost random}
\begin{algorithmic}[1]
\FORALL{cells} 
\STATE $c_i(t) = random[1-\Delta \cdots 1+ \Delta]$ 
\ENDFOR
\end{algorithmic}
\end{algorithm}
Alg. \ref{alg cost random walk} applies a random walk to each grid cell, but has no interaction between grid cells. As a result, we get a temporal correlation of the costs of each grid cell (Fig.~\ref{figure: cost time series}), but a spatially random pattern (Fig.~\ref{figure: cost surfaces}). To constrain the random walk around 1, an additional rebounding factor of $\omega \cdot \sqrt{|1-c_i(t-1)|}$ was added to the random walk. The scheduling within one time step is as follows:  
\begin{algorithm}[H]
\caption{Random Walk}\label{alg cost random walk}
\begin{algorithmic}[1]
\FORALL{cells} 
\STATE $c_i(t) = c_{i}(t-1) + \Delta  \cdot random[-1 \cdots1]$ \linebreak + $\omega \cdot sign(1-c_i(t-1)) \cdot \sqrt{|1-c_i(t-1)|}$ 
\ENDFOR
\end{algorithmic}
\end{algorithm}
Alg. \ref{alg cost correlated walk} applies a random walk with an additional spatial interaction to each grid cell. It produces spatio-temporally correlated costs (Fig.~\ref{figure: cost time series} and Fig.~\ref{figure: cost surfaces}). The scheduling within one time step is as follows:  
\begin{algorithm}[H]
\caption{Correlated Random Walk}\label{alg cost correlated walk}
\begin{algorithmic}[1]
\STATE Calculate average global costs
\FORALL{cells} 
\STATE$\bar{c_i} =$ average costs in the 8-cell neighborhood 
\STATE $c_i(t) = c_{i}(t-1) + \Delta  \cdot random[-1 \cdots 1] + \omega \cdot \bar{c_i} $
\STATE Normalize with average global costs 
\ENDFOR
\end{algorithmic}
\end{algorithm}

\section{Model Scheduling}\label{appendix: Scheduling}
\renewcommand{\theequation}{B.\arabic{equation}} 
\setcounter{equation}{0} 

\subsection{Economic Model}\label{appendix: economic scheduling}
The economic model is initialized with a random configuration which is at the desired cost level. To ensure that the random walks are in a steady state, we ran the simulation 10000 time steps before the ecological model was initialized. Each time step, the scheduling was as follows:
\begin{algorithm}[H]
\caption{Scheduling Economic Model}\label{alg scheduling economic model}
\begin{algorithmic}[1] 
\STATE Update costs
\REPEAT
\STATE{Adjust market price $P$}
\FORALL{cells} 
\IF{$P \cdot \xi_i(l) > c_i(t)$}
\STATE $x_i = 1$ (conserved) 
\ELSE
\STATE $x_i = 0$ (not conserved) 
\ENDIF
\ENDFOR
\STATE Calculate ecological value and costs
\UNTIL{budget constraint is met}
\STATE{Update land configuration}
\end{algorithmic}
\end{algorithm}

\subsection{Ecological Model}\label{appendix: ecologic scheduling}
The ecological model was started by randomly choosing $60\%$ of the patches as occupied. We checked that populations were in a steady state after initialization and thus the measurements were not affected by the initialization  \citep[see][]{Grimm-intrinsicmeantime-2004}. The scheduling of the ecological model within one time step is as follows:

\begin{algorithm}[H]
\caption{Scheduling Metapopulation Model}\label{alg scheduling ecological model}
\begin{algorithmic}[1]
\FORALL{populated cells} 
\STATE{Local extinction with rate $e$}
\ENDFOR
\FORALL{populated cells} 
\IF{Patch destroyed}
\STATE Disperse with emigration rate $r_d$
\ELSE
\STATE Disperse with emigration rate $r$
\ENDIF
\ENDFOR
\FORALL{unpopulated cells} 
\STATE Check if immigration succesfull
\ENDFOR 
\end{algorithmic}
\end{algorithm}

\end{appendices}

\end{document}